\documentclass[%
reprint,
superscriptaddress,
amsmath,amssymb,
aps,
twocolumn
]{revtex4-2}

\usepackage{amsmath}
\let\oldAA\AA
\renewcommand{\AA}{\text{\normalfont\oldAA}}
\usepackage{xcolor}
\usepackage{multirow}

\usepackage{booktabs}
\usepackage{lipsum}
\usepackage{graphicx}
\usepackage{dcolumn}
\usepackage{bm}
\usepackage{amsmath}
\usepackage{amssymb}
\usepackage{dutchcal}
\usepackage{float}
\usepackage{upquote}
\begin{document}

\title{How concerted are ionic hops in inorganic solid-state electrolytes?}

\author{Cibrán López}
    \affiliation{Departament de Física, Universitat Politècnica de Catalunya, 08034 Barcelona, Spain}
    \affiliation{Barcelona Research Center in Multiscale Science and Engineering, Universitat Politècnica de Catalunya, 08019 Barcelona, Spain}
    \affiliation{Institut de Ci\`encia de Materials de Barcelona, ICMAB--CSIC, Campus UAB, 08193 Bellaterra, Spain}

\author{Riccardo Rurali}
    \affiliation{Institut de Ci\`encia de Materials de Barcelona, ICMAB--CSIC, Campus UAB, 08193 Bellaterra, Spain}

\author{Claudio Cazorla}
    \affiliation{Departament de Física, Universitat Politècnica de Catalunya, 08034 Barcelona, Spain}
    \affiliation{Barcelona Research Center in Multiscale Science and Engineering, Universitat Politècnica de Catalunya, 08019 Barcelona, Spain}

\begin{abstract}
	Despite being fundamental to the understanding of solid-state electrolytes (SSE), little is known 
	on the degree of coordination between mobile ions in diffusive events. Thus far, identification 
	of concerted ionic hops mostly has relied on the analysis of spatio-temporal pair correlation functions 
	obtained from atomistic molecular dynamics (MD) simulations. However, this type of analysis neither 
	allows for quantifying particle correlations beyond two body nor determining concerted ionic hop 
	mechanisms, thus hindering a detailed comprehension and possible rational design of SSE. Here, we 
	introduce an unsupervised k-means clustering approach able to identify ion-hopping events and correlations 
	between many mobile ions, and apply it to a comprehensive \textit{ab initio} MD database comprising 
	several families of inorganic SSE and millions of ionic configurations. It is found that despite 
	two-body interactions between mobile ions are largest, higher-order $n$-ion ($2 < n$) correlations 
	are most frequent. Specifically, we prove an universal exponential decaying law for the probability 
	density function governing the number of concerted mobile ions. For the particular case of Li-based SSE, 
	it is shown that the average number of correlated mobile ions amounts to $10 \pm 5$ and that this 
	result is practically independent of temperature. Interestingly, our data-driven analysis reveals that 
	fast-ion diffusion strongly and positively correlates with ample hopping lengths and long hopping spans 
	but not with high hopping frequencies and short interstitial residence times. Finally, it is shown that 
	neglection of many-ion correlations generally leads to a modest overestimation of the hopping frequency 
	that roughly is proportional to the average number of correlated mobile ions. 
\end{abstract}

\maketitle

\section*{Introduction}
\label{sec:intro}
Solid-state electrolytes (SSE) presenting high ionic conductivity are pivotal for the development of 
transformative green-energy conversion and storage technologies like fuel cells, electrocatalysts and 
solid-state batteries \cite{famprikis19,sapkota20,mofarah19,aznar17}. SSE are complex materials that 
exhibit very disparate compositions, structures, thermal behaviors and ionic mobilities hence, unfortunately, 
it is difficult to rationally ascribe them to general categories and design principles \cite{bachman16,lopez23}. 
In particular, there is a lack of fundamental knowledge on the collective atomistic mechanisms that 
govern ionic transport. 

In recent years, analysis of the correlations between ionic transport (i.e., mobile ions) and lattice 
dynamics (i.e., vibrating ions) have attracted increasing interest \cite{lopez23,muy21,cazorla19,muy18}. 
The ``paddle-wheel'' mechanism, in which the libration of semirigid anionic units can propel cation 
transport \cite{zhang22}, is a well-known example of such a possible type of atomic concertation 
in superionic materials. The influence of lattice anharmonicity on ionic transport has been also thoroughly 
discussed, both theoretically and experimentally \cite{gupta21,ding20,ren23,xu22}. Nonetheless, very 
little is known on the existing level of coordination between many mobile ions in diffusive events. 

Thus far, identification of correlations between mobile ions mostly has relied on the analysis of van 
Hove correlation functions obtained from \textit{ab initio} molecular dynamics (AIMD) simulations and 
on zero-temperature nudged elastic band (NEB) calculations \cite{zhang19,jalem21,he17}. For Li-based SSE, 
it has been theoretically demonstrated that concertation between many mobile ions tends to lower the energy 
barriers for ionic diffusion, hence collective diffusive behaviour, rather than individual ionic hops, is 
expected to be predominant in superionic materials \cite{he17}. 

Nevertheless, due to the inherent limitations of the analysis methods employed up to now, many questions 
on the exact level of concertation between many mobile ions remain unanswered. For example, how many ions are 
typically coordinated in diffusive events and through which collective mechanisms? Are these many-ions correlations 
dependent on temperature or not? Can collective hopping behaviour be analytically described by a general law? 
Does the degree of ionic coordination depend on the specific family of SSE or is universal? How the neglection 
of many-ion correlation affects the estimation of key atomistic quantities like the ion hopping frequency? 
Answering these questions is not only relevant from a fundamental point of view, it is also necessary to justify 
the broad adoption of formulas obtained in the dilute-solution limit (e.g., the Nerst-Einstein relation for the 
ionic conductivity) which assume mobile ions to be fully uncorrelated \cite{ceder01,molinari21,marcolongo17,sasaki23,grossman19}.

\begin{figure*}[t]
\centering
\includegraphics[width=1.0\textwidth]{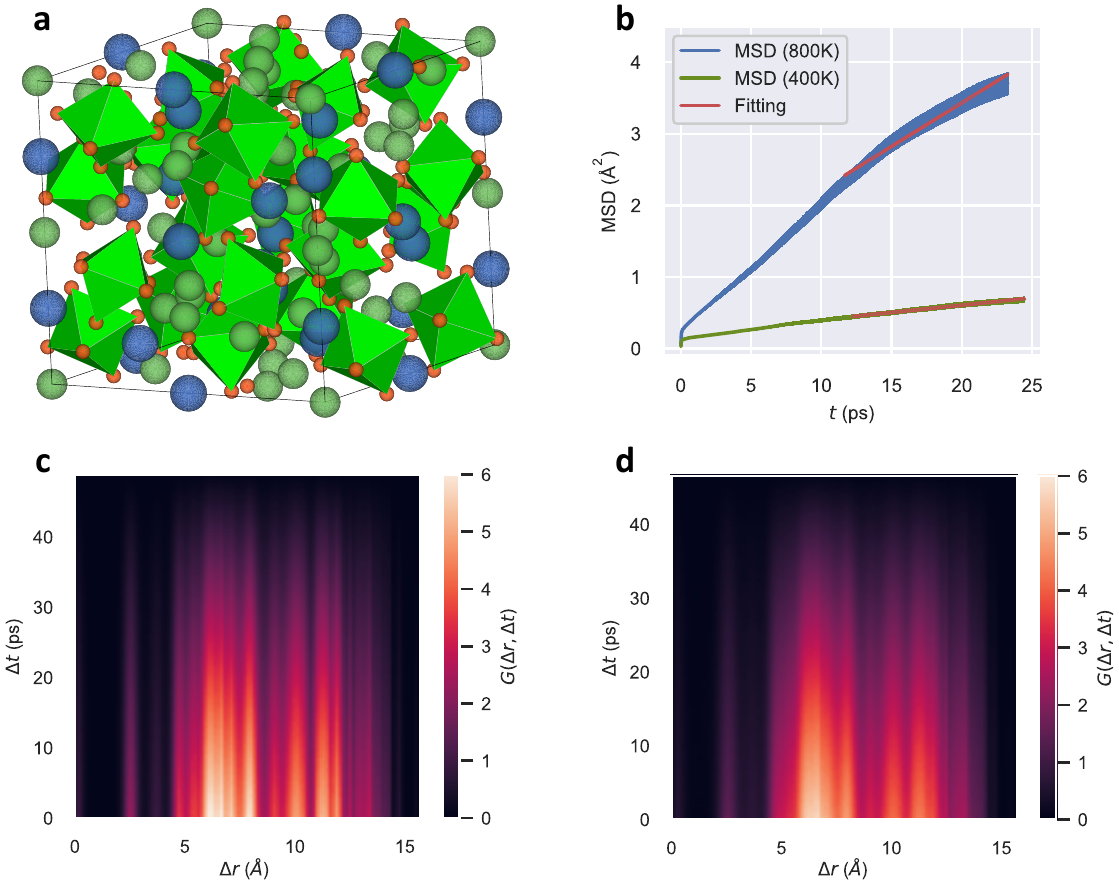}
	\caption{\textbf{Standard characterization of ionic transport and correlations in SSE from molecular
	dynamics simulations particularized for Li$_{7}$La$_{3}$Zr$_{2}$O$_{12}$ (LLZO).} (a)~Ball-stick 
	representation of bulk LLZO with a tetragonal crystal structure and space group $I4_{1}/acd$; lanthanum, 
	lithium, oxygen and zirconium atoms are represented with blue, green, red and blue spheres, respectively. 
	(b)~Mean squared displacement of Li cations obtained from DFT-AIMD simulations performed at $T = 400$ 
	and $800$~K. (c)--(d)~Van Hove correlation function for Li cations (in arbitrary units) obtained from 
	DFT-AIMD simulations performed at $T = 400$ and $800$~K, respectively.} 
\label{fig1}
\end{figure*}

In this work, we introduce a k-means clustering approach able to unsupervisedly identify ion-hopping events 
and quantify correlations between many mobile ions from ionic configurations generated in atomistic molecular 
dynamics simulations. This automatised analysis was recursively applied on a comprehensive AIMD database 
comprising several families of inorganic SSE and millions of atomic configurations \cite{lopez23,database}. It
was found that many-ion correlations beyond pairwise are dominant in diffusive events and can be represented by 
an universal exponential decaying law. Interestingly, for Li-based SSE it was determined that the average number 
of concerted mobile ions amounts to $10 \pm 5$, very much independently of temperature. Moreover, the introduced 
unsupervised analysis also permitted us to accurately quantify the prevalent correlations between ionic diffusion 
and key microscopic quantities like ion hopping lengths and frequencies and interstitial residence times. In addition, 
the effects of neglecting many-ion correlations on the estimation of the ion hopping frequency and migration energy
barrier were substantiated. Therefore, the present work leverages our fundamental understanding of technologically 
relevant SSE and elaborates on the adequacy of employing formulas obtained within the dilute-solution limit for 
describing them.

\section*{Results}
\label{sec:results}
Figure~\ref{fig1} shows the results of finite-temperature AIMD simulations performed for Li$_{7}$La$_{3}$Zr$_{2}$O$_{12}$ 
(LLZO), an archetypal Li-based SSE \cite{murugan07}. LLZO is a complex oxide material with garnet-like structure (space 
group $I4_{1}/acd$, Fig.~\ref{fig1}a) that presents high lithium-ion conductivity and excellent thermal and chemical 
stabilities. As it is customarily done for SSE, one can estimate the tracer Li diffusion coefficient of LLZO, $D_{\rm Li}$,  
directly from the configurations generated during AIMD simulations by computing the time derivative of the corresponding 
mean squared displacement (Fig.~\ref{fig1}b and Methods) \cite{cazorla19,islam21}. Larger $D_{\rm Li}$ values are associated 
with larger ionic conductivities, $\sigma_{\rm Li}$, as deduced from the popular Nernst-Einstein relation obtained in the
dilute-solution limit:
\begin{equation}
\sigma_{\rm Li} = \frac{z_{\rm Li} F}{k_{B}T} \cdot D_{\rm Li}~,
\label{eq:nerst}
\end{equation}
where $z_{\rm Li}$ represents the charge of the mobile ion, $k_{B}$ the Boltzmann constant and $F = e \cdot N_{\rm A} $ 
the Faraday constant ($e$ is the electron charge and $N_{\rm A}$ the Avogadro's number). 

\begin{figure*}[t]
\centering
\includegraphics[width=1.0\textwidth]{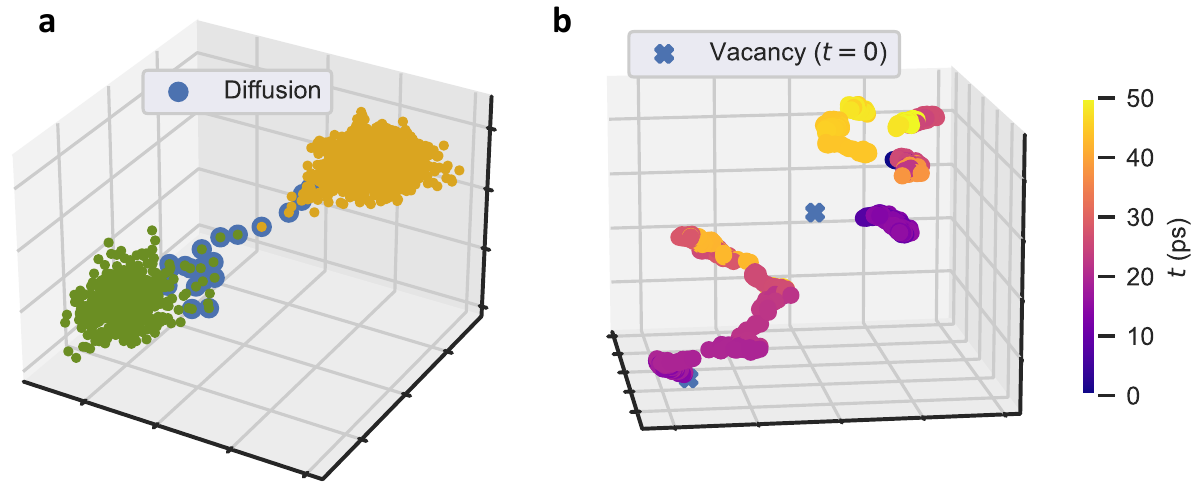}
        \caption{\textbf{Unsupervised k-means clustering algorithm for identification of ionic hops and diffusion paths.} 
		(a)~Ionic diffusion of an arbitrary mobile atom in a DFT-AIMD simulation of LLZO performed at $T = 400$~K 
		(blue circles). The two vibration centers defining the origin and end of the ionic hop are represented 
		with orange and green points, respectively. (b)~Temporal sequence of ionic hops identified for a $\approx 50$~ps 
		duration DFT-AIMD simulation of LLZO performed at $T = 400$~K. Blue crosses represent the initial position 
		of two lithium vacancies introduced in the simulation cell; ionic hops are initiated near them. Different 
		sections of a same diffusion path do not necessarily correspond to a same ion.}
\label{fig2}
\end{figure*}

The van Hove correlation function, $G (\Delta r, \Delta t)$ (Methods), provides information on the spatio-temporal distribution
of pairs of particles in atomistic configurations obtained from finite-temperature MD simulations (e.g., for a null time
span $G$ is equivalent to the usual radial pair distribution function). Figures~\ref{fig1}c,d show the van Hove correlation
function of Li atoms estimated for superionic LLZO at two different temperatures; it is appreciated that pair correlations
between nearby ions (i.e., $2 \le \Delta r \le 5$~\AA) are substantial over time spans of several tens of picoseconds, since 
$G (\Delta r, \Delta t)$ remains discernible within those variable intervals. At the highest simulated temperature, ionic 
diffusion is sizeable (Fig.~\ref{fig1}b) and the peaks of the van Hove correlation function (Fig.~\ref{fig1}d) get noticeably 
faded (barely change) along the interparticle distance (time) dimension in comparison to those obtained for the non-superionic 
state (Fig.~\ref{fig1}c). For completitude purposes, the ``self'' and ``distinct'' components of the lithium van Hove correlation 
function (Methods) are shown in Supplementary Fig.1. These $G (\Delta r, \Delta t)$ results clearly show the existence of 
significant ion-pair correlations in LLZO ionic diffusion.

Nevertheless, the standard particle correlations analysis presented above is too restricted since it only considers correlations
between pairs of atoms, thus neglecting any possible higher-order level of $n$-ion ($2 < n$) concertation. In addition, it does
not provide any atomistic insight into the many-ion mechanisms involved in ionic diffusion. To overcome this type of limitations, 
we devised an algorithm based on k-means clustering that is able to unsupervisedly identify ion-hopping events and correlations
between many particles, and applied it to a comprehensive AIMD database of inorganic SSE \cite{lopez23,database}. The introduced 
algorithm also permits to automatically identify ion hopping lengths and frequencies and interstitial residence times, hence 
the general dependencies between these atomistic descriptors and ionic diffusion can be determined.

\subsection*{K-means clustering algorithm for unsupervised identification of ionic hops and diffusive paths}
\label{subsec:k-means}
Our approach consists in identifying the equilibrium and metastable positions in a supercell around which particles vibrate; 
subsequently, the temporal sequence of atomic displacements from one of those vibrational centers to another are monitored 
thus determining ion diffusion paths without imposing any restriction. Only two fundamental premises are assumed in our 
procedure, namely, the vibration of ions around equilibrium and metastable positions are roughly isotropic, and diffusion 
events are less frequent than atomic vibrations.

K-means clustering is an unsupervised machine learning algorithm that classifies objects in such a way that elements 
within a same group, called ``cluster'', are in a broad sense more similar to each other than to elements in other clusters.
Our method for identifying vibrational centres from sequential ionic configurations relies on k-means clustering (Methods) 
since this approach assumes isotropy on the fluctuations of non-diffusive particles. (It is worth mentioning that spectral 
clustering, based on interparticle connectivity instead of interparticle distance, was also considered, however less 
satisfactory ionic hops identification results were obtained in this case.) Importantly, the definition of arbitrary 
materials-dependent threshold distances for scrutiny of ionic hops is completely avoided in our approach, as we explain next. 

For each individual ionic trajectory, the optimal number of clusters, $K$, which represents the number of vibrational 
centres that the particle visits during the simulation, is systematically selected as the one that maximises the 
silhouette coefficient averaged over all the samples corresponding to cases $2 \le K$ (Methods). Silhouette coefficients, 
$S$, are individually ascribed to each cluster and can take values within the interval $[-1,+1]$. $S$ values near $+1$ indicate 
that the sample is far away from the neighbouring clusters. On the other hand, negative $S$ values indicate that the sample might 
has been assigned the wrong cluster (an exact zero value would indicate that the sample is on the decision boundary between 
two neighbouring clusters). Nevertheless, this procedure fails to describe the case of a non-diffusive particle, which would 
correspond to $K = 1$, since by construction $2 \le K$. To avoid this issue, whenever the maximum average silhouette coefficient 
is below an arbitrary, but reasonable, threshold value of $0.7$, we automatically impose $K = 1$ (i.e., the ion does not diffuse 
throughout the simulation). The dependence of our algorithm performance on such a threshold value has been exhaustively tested, 
finding negligible effects on the final outcomes.

\begin{figure*}[t]
\centering
\includegraphics[width=1.0\textwidth]{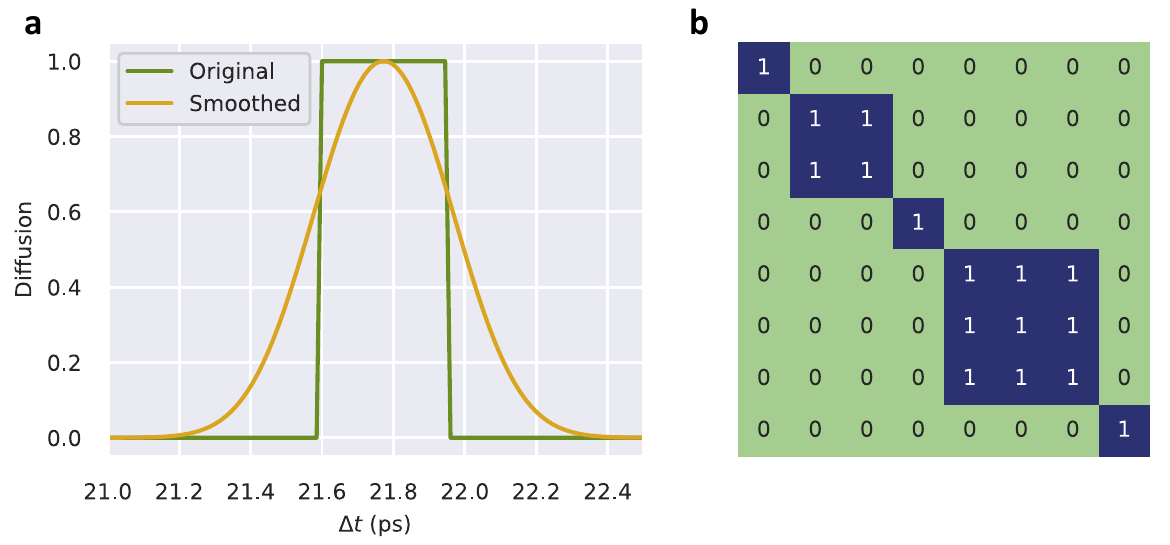}
        \caption{\textbf{Unsupervised estimation of correlations between many mobile ions.} 
	(a)~At each time step, the state of each mobile ion is identified with a ``0'', if it is vibrating, or a ``1'', if 
	it is hopping. The multi-step functions obtained over time are smoothed out with Gaussian functions to improve the 
	numerical convergence in the subsequent calculation of the many-ion correlation matrix. (b)~Considering all the binary 
	data generated during a molecular dynamics simulation, a $N \times N$ correlation matrix is obtained, $N$ being the 
	number of possible mobile ions, which provides the number and indexes of uncorrelated and correlated ions (represented 
	by ``0'' and ``1'', respectively). In the provided example, a group of two ions and another of three move concertedly, 
	while three particles remain uncorrelated during the whole simulation.} 
\label{fig3}
\end{figure*}

Once the number of vibrational centres, their real-space location and temporal evolution are determined, ionic diffusive 
paths are defined like the sections connecting two different vibrational centres over time. In our calculations, it is 
appreciated that, due to the discrete nature of the generated ionic trajectories, diffusive paths generally start and finish 
at around $0.5$~\AA~ from their defining vibrational centres. An illustrative example of our method for identification of 
vibrational centres and ionic diffusive paths is shown in Fig.~\ref{fig2}a. Therein, two vibrational centres with a highly 
confident average silhouette coefficient value of $0.88$ (green and yellow points) are depicted along with the ionic diffusive 
path (blue points) that connects them. Our algorithm was recursively applied to a comprehensive DFT-AIMD database involving 
different families of SSE \cite{lopez23,database} (Supplementary Tables~1--3), obtaining in all the cases highly accurate 
results for the identification of ionic hops and diffusive paths. For example, for non-stoichiometric LLZO (i.e., containing 
Li vacancies) simulated at temperatures of $400$ and $800$~K, reassuring average silhouette coefficients amounting to 
$0.99$ and $0.97$ were respectively obtained (Fig.~\ref{fig2}b).

It is worth noting that our ionic hop identification algorithm neither presupposes a fixed number nor the positions of 
the vibrational centres in the provided atomistic configurations (e.g., the number of vibrational centres may differ 
from the number of potentially mobile atoms when there is significant ionic diffusion). This adaptability feature turns 
out to be particularly useful for the identification of metastable crystalline positions (e.g., interstitials) and 
evaluation of residence times, as we will show later on. The analysis method just explained has been implemented in the 
\verb!IonDiff! software \cite{iondiff}, a freely available open-source python code (Methods).

\subsection*{Quantitative analysis of concertation between many mobile ions}
\label{subsec:correl}
The ionic hop identification approach explained above was applied to a comprehensive DFT database of inorganic SSE 
comprising a total of $83$ AIMD simulations (Methods) in which ionic diffusion was substantial \cite{lopez23,database}. 
Since we are primarily interested in unveiling universal behaviours and relationships in ionic transport, we considered 
different Ag-, Cu-, halide-, O-, Na- and Li-based superionic compounds (Supplementary Tables~1--3). To quantitatively 
evaluate the correlations and level of concertation between an arbitrary number of mobile ions, $n$, we devised and 
implemented the following algorithm.

For a given sequence of ionic configurations generated during a molecular dynamics simulation, the corresponding 
correlation matrix for diffusive events was computed. To this end, we first assigned a value of ``1'' to each diffusing 
particle and of ``0'' to each vibrating particle at each simulated time frame (Fig.~\ref{fig3}a). Such a binary 
numerical assignment was straightforwardly performed with the ionic hop identification algorithm introduced in the 
previous section. Due to the discrete nature of the generated ionic trajectories, and to improve numerical convergence 
in the subsequent correlation analysis, the obtained multi-step time functions were approximated with Gaussians that 
equaled the half maxima at their width (Fig.~\ref{fig3}a, in analogy to the ``full-width-at-half-maximum'' --FWHM-- method 
widely employed in signal processing). Subsequently, we computed the $N \times N$ correlation matrix, where $N$ is 
the number of potentially mobile ions, resulting from all the gathered simulation data; this latter step involves the 
calculation of covariance coefficients for ions taken in pairs \cite{lopez23}.  

\begin{figure*}[t]
\centering
\includegraphics[width=1.0\textwidth]{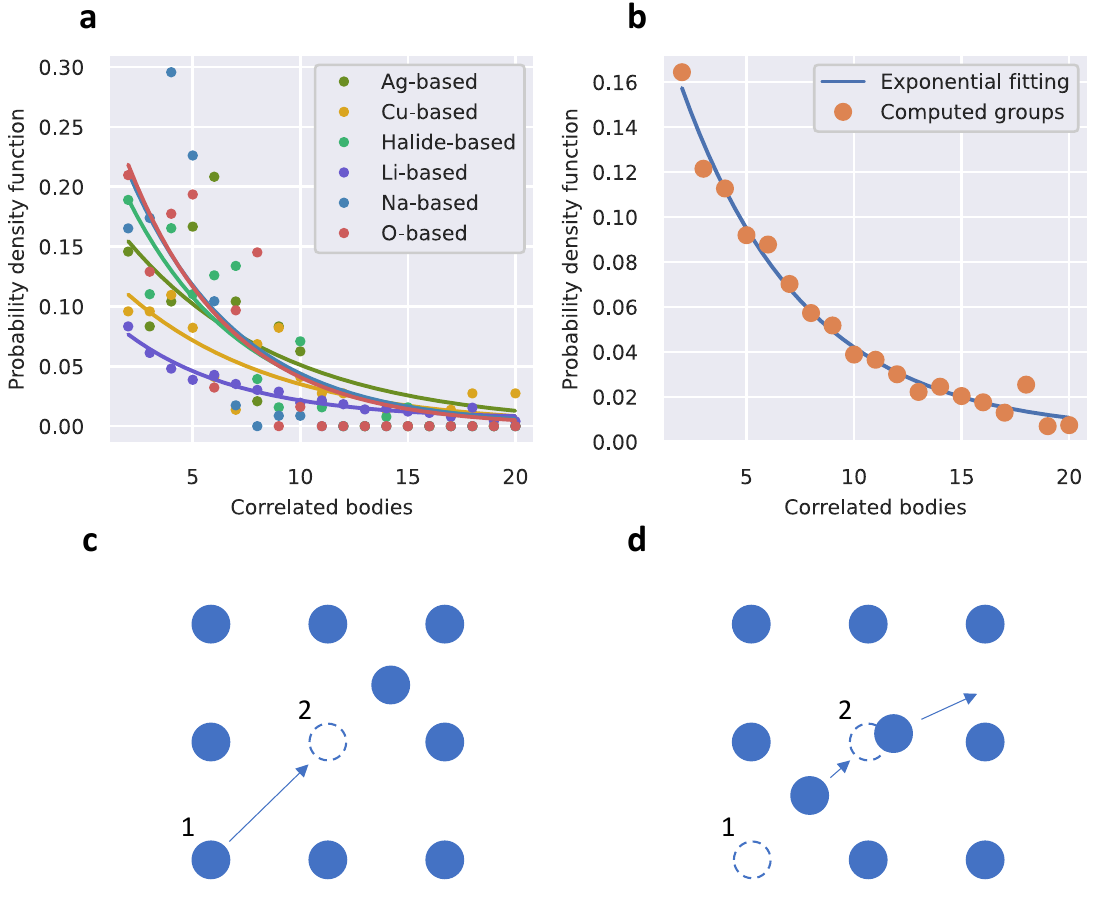}
        \caption{\textbf{Many-ion correlation results obtained from applying the introduced unsupervised k-means clustering 
	algorithm on a comprehensive DFT-AIMD database of inorganic SSE \cite{database}.} 
	(a)~Probability distribution function for the number of coordinated ions in collective ionic hops estimated separately 
	for each SSE family. High-order many-ion correlations are most substantial in Cu- and Li-based superionic materials. 
	Solid lines represent exponential decaying fits to the data points. (b)~General probability distribution function for 
	the number of coordinated ions in collective ionic hops estimated considering all SSE compounds. An exponential decaying 
	function of the form $f(n) = 0.220 \cdot \exp{\left( -0.252n \right)}$ fits fairly well the obtained data points. 
	Two different two-ion coordinated mechanisms were most frequently observed in diffusive events: (c)~ion (1) moves towards 
	an empty equilibrium lattice position just left by ion (2), and (d)~a mobile ion (1) kicks out a vibrating atom (2) and 
	occupies its equilibrium lattice position.}
\label{fig4}
\end{figure*}

The correlation matrix thus estimated, however, may be difficult to converge due to its statistical character (particularly
in situations for which the number of mobile ions and time steps are somewhat limited, as it tends to be the case of 
computationally intensive AIMD simulations). To overcome these practical issues, we numerically computed a reference 
correlation matrix corresponding to a randomly-distributed sequence of ionic hops with Gaussian FWHM equal to the mean 
diffusion time determined for the scrutinised simulation (note that due to the finite width of the Gaussians such a 
correlation matrix is not exactly equal to the identity). Subsequently, covariance coefficients in the original 
correlation matrix larger (smaller) than the corresponding random reference values were considered as true correlations 
(random noise), hence were rounded off to one (zero) for simplification purposes. In order to not underrate the many-ion 
correlations, different hops of a same ion were treated as independent events. 

In this manner, a correlation matrix consisting of ones and zeros is finally assembled from which one can easily determine 
how many and which particles remain concerted during diffusion. Figure~\ref{fig3}b, shows a correlation matrix example in 
which a group of two mobile atoms and another of three move concertedly, while three ions remain uncorrelated during the whole 
simulation (rows and columns have been reshuffled in order to facilitate the visualisation of many-ion correlations). The 
described many-ion correlation identification algorithm also has been implemented in the \verb!IonDiff! software \cite{iondiff}, 
a freely available open-source python code (Methods).

\subsection*{Probability density function governing the number of correlated mobile ions}
\label{subsec:unipdf}
Figure~\ref{fig4}a shows the probability density function (pdf) that governs the number of concerted ions in diffusive 
events estimated for different SSE families (i.e., averaged over compounds belonging to a same category and temperature). 
These results were obtained from AIMD simulations that fully take into account anharmonicity and temperature effects. 

In all the cases, an exponential decaying function was found to fairly reproduce the estimated distribution of $n$-concerted 
ions (solid lines in Fig.~\ref{fig4}a). Consequently, the degree of concertation between mobile particles is always largest 
for pairs of ions and steadily decreases for increasing number of ions (here, we arbitrarily but reasonably considered 
only cases up to $n = 20$). The value of the pre-exponential factor and parameter in the exponential function, 
however, significantly vary from one family of materials to another. Therefore, the level of many-ion coordination in 
diffusive events depends on the specific SSE group. In particular, O-, halide- and Na-based fast-ion conductors exhibit 
the most rapidly decaying pdf profiles, meaning that correlations for a large number of mobile ions are smallest. On the 
other hand, Cu- and Li-based fast-ion conductors display the most slowly decaying pdf profiles (i.e., correlations for 
a large number of mobile ions are largest), while Ag-based SSE render an intermediate trend. 

Figure~\ref{fig4}b shows the general probability density function obtained for the number of concerted mobile ions in 
fast-ion conductors (i.e., averaged over all SSE families and temperature). An exponential decaying law is found to 
reproduce remarkably well the estimated distribution of $n$-ion correlations. In this general case, the degree of particles 
concertation is also largest for pairs of ions, as expected. However, by performing integrations of the area enclosed 
below the solid line in Fig.~\ref{fig4}b, it is found that coordinated diffusive events involving more than two ions turn 
out to be more frequent (roughly by a factor of $6$). This finding, which follows from comprehensive AIMD simulations and 
is not restricted to an unique SSE family, is consistent with previous computational results reported for Li-based 
materials \cite{he17}. 

\begin{figure*}[t]
\centering
\includegraphics[width=1.0\textwidth]{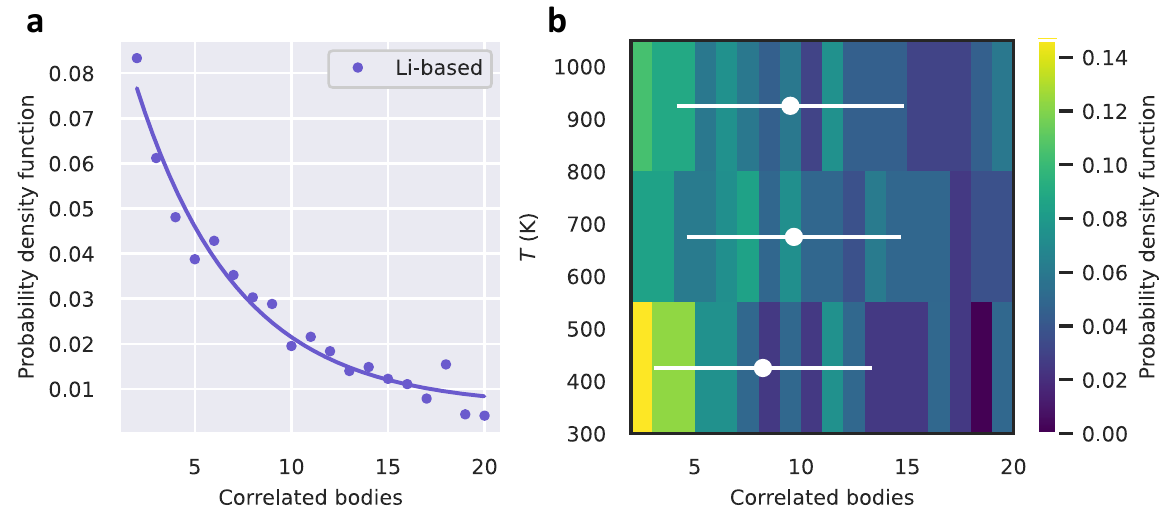}
        \caption{\textbf{Many-ion correlation results obtained for Li-based SSE.} 
	(a)~Probability distribution function for the number of concerted ions in diffusive events.
	The solid line represents an exponential decaying fit to the data points.
	(b)~Temperature-dependence of the probability distribution function shown in (a) considering
	the intervals $300 \le T_{1} \le 550$~K, $550 \le T_{2} \le 800$~K and $800 \le T_{3} \le 1050$~K. 
	White dots and lines denote average values and corresponding standard deviations, respectively.}
\label{fig5}
\end{figure*}

Our formalism also allows to identify which particles participate in the disclosed $n$-ion coordinated diffusion 
events, which is very convenient for data visualization purposes. For the special case of $n=2$ correlated 
diffusion processes, we determined the two most relevant atomistic coordination mechanisms, which are sketched 
in Figs.~\ref{fig4}c,d. The first mechanism consists in a sequence of two diffusion events in which a first 
mobile ion hops to an interstitial position leaving a vacant site that is immediately occupied afterwards by a 
second diffusing particle (Fig.~\ref{fig4}c). The second mechanism consists in the forced jump of a particle 
resulting from the direct influence of a second diffusing ion (Fig.~\ref{fig4}d). It is worth noting that these 
two $n = 2$ ionic correlation mechanisms have been already reported in the literature for Li-based compounds 
\cite{he19}, thus confirming the reliability of our unsupervised ionic-hop identification approach.

\subsection*{Temperature dependence of many mobile ions correlations}
\label{subsec:tdepen}
An interesting question to answer for superionic materials is whether the degree of concertation between many mobile 
ions depends on temperature or not \cite{ceder01,molinari21,winter23}. The findings reported in the previous section 
cannot provide direct insights into this question since were obtained from thermal averages. Consequently, we performed 
a detailed temperature analysis of the many-ion correlations identified for Li-based compounds alone, since these are 
technologically very relevant and relatively abundant.  

Figure~\ref{fig5}a shows the pdf estimated for the number of concerted many mobile ions in Li-based SSE (same as in 
Fig.~\ref{fig4}a). By taking all the collective diffusive events represented in that figure, we constructed normalised 
temperature histograms considering the three intervals $300 \le T_{1} \le 550$~K, $550 \le T_{2} \le 800$~K and $800 
\le T_{3} \le 1050$~K, as shown in Fig.~\ref{fig5}b. Very mild differences are appreciated for the pdf's estimated 
for such temperature ranges. For example, at low temperatures coordinated diffusion events involving pairs of ions 
appear to be more frequent than at high temperatures. However, when average quantities are considered, such moderate 
discrepancies mostly disappear. Specifically, the average number of coordinated mobile ions approximately amounts to 
$10 \pm 5$ for all the investigated temperature intervals (white dots and lines in Fig.~\ref{fig5}b). Therefore, we may 
conclude that the level of concertation between mobile ions in Li-based SSE is practically independent of temperature.

\subsection*{Relationship between ionic diffusion and key atomistic descriptors}
\label{sec:descriptors}
As explained in previous sections, the \verb!IonDiff! software \cite{iondiff} allows to determine the centers of 
vibration and exact migrating paths of ions as provided by molecular dynamics simulations. Accordingly, for a given 
sequence of ionic configurations, it is straightforward to estimate insightful atomistic descriptors like the average 
hopping distance, $\Delta r$, hopping time, $\Delta t$, and hopping frequency, $\nu$. Likewise, it is also 
possible to estimate interstitial residence times, $\gamma$, by monitoring the simulation time during which a particle 
remains fluctuating around a metastable position (e.g., interstice). The identification of metastable positions
was performed by comparing the centers of vibration obtained during a whole simulation with those of the perfect 
equilibrium configuration, and assuming that metastable and equilibrium vibrational centers should be separated by 
a distance of at least $1.0$~\AA~ (Supplementary Fig.2).

\begin{figure*}[t]
\centering
\includegraphics[width=1.0\textwidth]{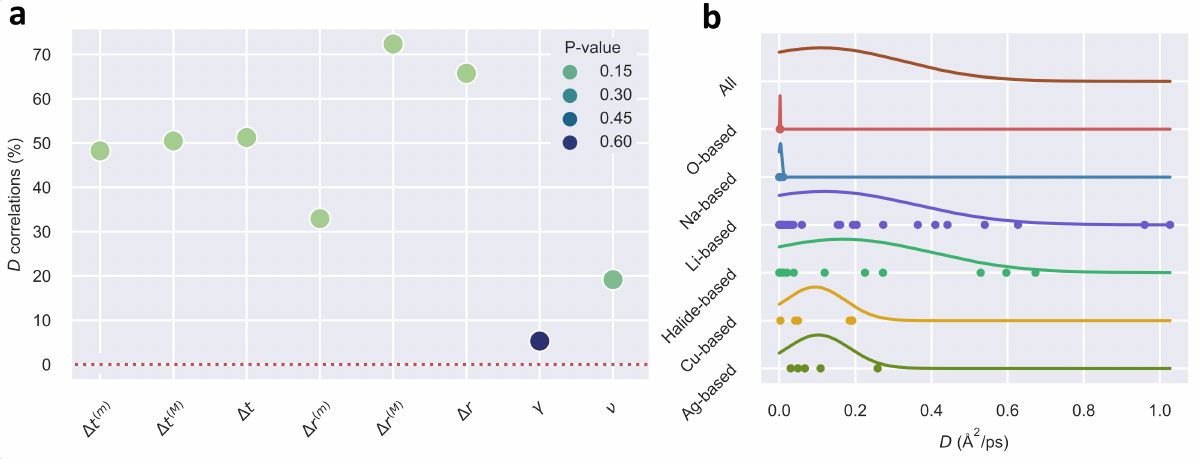}
        \caption{\textbf{Correlations between ionic diffusion and key atomistic descriptors.} 
	(a)~$D$ stands for the tracer ion diffusion coefficient, $\Delta t$ the average duration of an ionic hop, 
	$\Delta r$ the average length of an ionic hop, $\gamma$ the average interstitial residence time and $\nu$ 
	the hopping frequency. Superscripts ``(M)'' and ``(m)'' denote maximum and minimum values estimated for the 
	corresponding descriptor. (b)~Distribution of tracer ion diffusion coefficients calculated for each SSE 
	family.}
\label{fig6}
\end{figure*}

Figure~\ref{fig6} shows the level of correlation estimated for the tracer ion diffusion coefficient, $D_{\rm x}$, and 
atomistic descriptors described above considering all the SSE families examined in this study (for this analysis,
we considered the tracer diffusion coefficient instead of the full ion diffusion coefficient 
\cite{molinari21,marcolongo17,sasaki23} because of its ubiquity in computational studies). Such correlations were obtained 
by following the same data-analysis approach that was introduced in work \cite{lopez23}, which essentially involves the 
computation of Spearman correlation coefficients and $p$-values for the assessment of statistical significance. Besides 
examining average quantities, for the case of $\Delta r$ and $\Delta t$ we also considered their maximum, ``(M)'', and 
minimum, ``(m)'', values. Several interesting conclusions follow from the results shown in Fig.~\ref{fig6}a.

The largest $D_{\rm x}$ correlations involving average quantities are found for the hopping length and hopping time, 
which are both positive and roughly amount to $65$ and $50$\%, respectively. In the particular case of $\Delta r$, 
the maximum ion diffusion correlation is obtained for its maximum value, $\Delta r^{(M)}$, which is above $70$\% 
(Fig.~\ref{fig6}a). On the other hand, the smallest $D_{\rm x}$ correlation is found for the average interstitial 
residence time, which only amounts to $\approx 5$\%. As for the hopping frequency, the level of correlation with 
the ion diffusion coefficient is also positive but quite reduced ($\approx 20$\%). In most cases, the estimated 
correlations turn out to be statistically significant since the accompanying $p$-values are equal or smaller than 
$0.10$ \cite{lopez23}. For a detailed description of the examined data, Fig.~\ref{fig6}b shows the distribution of 
tracer ion diffusion coefficients calculated for each SSE family, which turn out to be quite diverse. 

Based on this data-driven atomistic analysis, we may conclude that good superionic materials characterized by large ion 
diffusion coefficients should present large hopping lengths and hopping times but not necessarily high hopping frequencies
and/or short interstitial residence times (Fig.~\ref{fig6}a). To put it differently, ample and timely, rather than short 
and too frequent, ionic hops appear to be associated with high ionic diffusion. 

To gain further insight into the connections between high ionic diffusion and key atomistic descriptors, Supplementary Fig.3
shows the $T$-dependence of $\nu$ and $\Delta r$ as evaluated for different SSE families. In general, it is found that the 
hopping frequency does not appreciably change with temperature whereas the average hopping distance noticeably increases 
upon increasing temperature. These results imply that the general $T$-induced ionic diffusion enhancement observed in SSE 
mostly is mediated by a surge in $\Delta r$ rather than in $\nu$. In turn, these findings appear to be coherent with the 
main conclusion presented in the preceding paragraph, namely, that the influence of the average hopping distance on fast-ion 
conduction exceeds that of the hopping frequency.

\section*{Discussion}
\label{sec:discussion}
In the dilute-solution limit, the interactions between mobile ions are regarded as negligible hence the full ionic diffusion 
coefficient reduces to the tracer diffusion coefficient \cite{molinari21,marcolongo17,sasaki23} (Methods) and its dependence 
on temperature can be expressed as \cite{cazorla19,ceder01}:
\begin{eqnarray}
	D_{\rm x} (T) & = & D_{{\rm x},0} \cdot \exp{{\left(-\frac{E_{a}}{k_{B}T} \right)}}  \nonumber \\ 
		      &   & D_{{\rm x},0} \propto a^{2} \nu_{0}~, 
\label{eq:diffusivity}
\end{eqnarray}
where $a$ is a hopping distance, $E_{a}$ the activation energy barrier for ionic migration and $\nu_{0}$ the hopping frequency. 

The many mobile ion correlation results presented in previous sections show that the dilute-solution limit in general 
does not apply to technologically relevant SSE, hence one may question the validity of Eq.(\ref{eq:diffusivity}) 
and other commonly employed formulas, like the Nerst-Einstein relation [Eq.(\ref{eq:nerst}) above], obtained under 
similar approximations. Aimed at quantitatively exploring this objection, we computed the hopping frequencies of all the 
SSE analysed in this study by using Eq.(\ref{eq:diffusivity}), $\nu_{0}$, which assumes the interactions between mobile 
ions to be negligible, and compared them with the values obtained directly from AIMD simulations with the \verb!IonDiff! 
software \cite{iondiff}, $\nu$, which fully takes into consideration many-ion correlations. Since an undetermined proportionality 
factor enters Eq.(\ref{eq:diffusivity}), we constrain our comparative analysis to the orders of magnitude of the examined 
hopping frequencies.

\begin{figure}[t]
\centering
\includegraphics[width=0.5\textwidth]{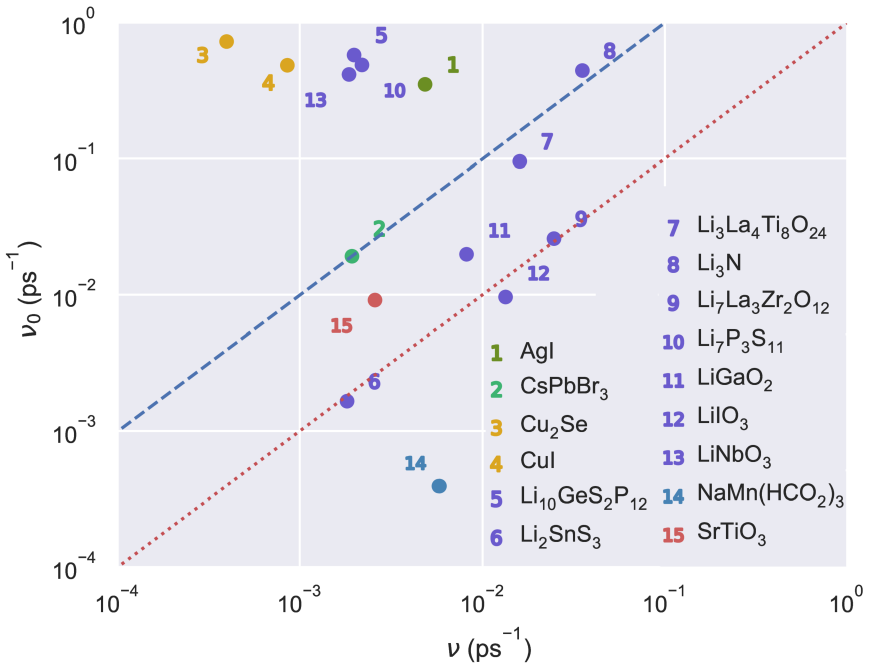}
        \caption{\textbf{Comparison of the hopping frequencies estimated for representative SSE in the dilute-solution limit, 
	$\nu_{0}$, and by explicitly considering many-ion correlations, $\nu$.}
	The straight lines in the plot indicate the \textit{coincidence} region in which the orders of magnitude of the
	two represented hopping frequencies coincide or differ to within a factor of $10$ while fulfilling the condition
	$\nu \le \nu_{0}$ (main text).}
\label{fig7}
\end{figure}

Figure~\ref{fig7} shows our $\nu_{0}$ and $\nu$ results obtained for $15$ representative superionic materials. Due to the 
fact that the proportionality factor entering Eq.(\ref{eq:diffusivity}) may be of the order of $10^{0}$--$10^{1}$, we regard 
as coincident a pair of $\nu_{0}$--$\nu$ hopping frequencies differing within such a quantity and fulfilling the condition 
$\nu \le \nu_{0}$ (i.e., the \textit{coincidence} region delimited by the straight lines $\nu_{0} = \nu$ --red-- and $\nu_{0} 
= 10 \nu$ --blue-- in Fig.~\ref{fig7}). It is appreciated that by neglecting many-ion correlations the hopping frequency
is slightly overestimated in average. In particular, $6$ out of the $15$ analysed materials are represented by points that 
clearly lie on the outer region above the selected coincidence interval. For instance, a large frequency discrepancy amounting 
from one to two orders of magnitude are obtained for Li$_{10}$GeS$_{2}$P$_{12}$, LiNbO$_{3}$, Cu$_{2}$Se, CuI and AgI. On the 
other hand, the $\nu$'s estimated for Li$_{7}$La$_{3}$Zr$_{2}$O$_{12}$, Li$_{2}$SnS$_{3}$, SrTiO$_{3}$ and CsPbBr$_{3}$, among 
others, agree fairly well with the approximate hopping frequencies obtained from the corresponding tracer diffusion coefficients.

For Li-, Cu- and Ag-based SSE, the results enclosed in Fig.~\ref{fig7} indicate that $\nu_{0}$ in general is a not a good 
approximation for $\nu$ since the former overestimates the latter. Contrarily, the points obtained for halide-, Na- and 
O-based SSE, as well as for some Li-based, are located inside or very close to the selected coincidence region meaning that 
$\nu_{0}$ is a reasonably good approximation for $\nu$. Based on these findings, along with those presented in previous 
sections (Fig.~\ref{fig4}a), we can state that the hopping frequency of materials in which the correlations between 
mobile particles extend to many ions (only few ions) are likely to be poorly (fairly well) approximated by the tracer 
diffusion coefficient. This conclusion is quantitatively novel since failure of the relations obtained in the dilute-solution 
limit now can be directly associated with the average number of correlated mobile ions.

Finally, in order to quantify the influence of neglecting many-ion correlation on the calculation of the activation energy 
barrier for ionic migration, $E_{a}$, we estimated this quantity for Li$_{7}$La$_{3}$Zr$_{2}$O$_{12}$ (LLZO) and 
Li$_{10}$GeS$_{2}$P$_{12}$ (LGSP) considering both the tracer and full ionic diffusion coefficients (Methods) 
\cite{molinari21,marcolongo17,sasaki23}. We selected these two materials because the first lies inside the coincidence 
interval defined for the ion hopping frequency while the second outside. For LLZO, it was found that when disregarding 
many-ion correlations $E_{a}$ amounted to $0.16$~eV whereas it decreased to $0.14$~eV when accounting for them. For LGSP, 
we obtained similar results, in particular, $0.21$ and $0.20$~eV from the tracer and full ionic diffusion coefficients, 
respectively. Therefore, it may be concluded that the influence of neglecting many-ion correlations on the estimation of 
$E_{a}$ appears to be less significant than for $\nu$.  
\\

In conclusion, we have carried out a comprehensive and unsupervised many mobile ion correlation analysis for several families 
of SSE based on the k-means clustering approach, which has been implemented in the freely available open-source python code 
\verb!IonDiff! \cite{iondiff}. An exponential decaying law is found to correctly describe the general probability density 
distribution governing the degree of concertation between many mobile ions in SSE. Accordingly, $n$-ion coordinated diffusion 
processes with $2 < n$ are found to be more frequent than pairwise coordinated diffusive events, although the latter hold the 
largest individual probability. For the particular case of Li-based SSE, the average number of correlated mobile ions is 
estimated to be $10 \pm 5$ and, interestingly, this result turns out to be practically independent of temperature. Furthermore, 
our data-driven analysis concludes that promising superionic materials characterized by large ion diffusion coefficients strongly 
and positively correlate with ample hopping lengths and long hopping times but not with high hopping frequencies and short 
interstitial residence times. Finally, it is shown that neglecting many-ion correlations generally leads to a modest 
overestimation of the hopping frequency that roughly is proportional to the average number of correlated mobile ions. Overall, 
our work leverages the fundamental understanding of ionic transport and superionic materials and elaborates on the limitations 
of using formulas obtained in the dilute-solution approximation for describing technologically relevant SSE. 
\\

\section*{Methods}
\label{sec:methods}
{\bf First-principles calculations outline.}~\textit{Ab initio} calculations based on density functional theory (DFT)
\cite{cazorla17} were performed to analyse the physico-chemical properties of bulk SSE. We performed these calculations 
with the VASP code \cite{vasp} by following the generalized gradient approximation to the exchange-correlation energy 
due to Perdew \emph{et al.} \cite{pbe96}. (For some halide compounds, likely dispersion interactions were captured 
with the D3 correction scheme developed by Grimme and co-workers \cite{grimmed3}.) The projector augmented-wave method 
was used to represent the ionic cores \cite{bloch94} and for each element the maximum possible number of valence 
electronic states was considered. Wave functions were represented in a plane-wave basis typically truncated at $750$~eV. 
By using these parameters and dense ${\bf k}$-point grids for Brillouin zone integration, the resulting zero-temperature 
energies were converged to within $1$~meV per formula unit. In the geometry relaxations, a tolerance of 
$0.005$~eV$\cdot$\AA$^{-1}$ was imposed in the atomic forces.
\\

{\bf First-principles molecular dynamics simulations.} \emph{Ab initio} molecular dynamics (AIMD) simulations based on DFT were 
performed in the canonical $(N,V,T)$ ensemble (i.e., constant number of particles, volume and temperature) for all the analysed 
materials. The selected volumes were those determined at zero temperature hence thermal expansion effects were neglected; 
nevertheless, based on previously reported molecular dynamics tests \cite{cazorla19}, thermal expansion effects are not expected 
to affect significantly the estimation of ion-transport features at moderate temperatures. The concentration of ion 
vacancies in the non-stoichiometric compounds was also considered independent of the temperature and equal to $\sim 1$--$2$\%. 
The temperature in the AIMD simulations was kept fluctuating around a set-point value by using Nose-Hoover thermostats. Large 
simulation boxes containing $N \sim 200$--$300$ atoms were employed in all the cases and periodic boundary conditions were 
applied along the three supercell vector directions. Newton's equations of motion were integrated by using the customary Verlet's 
algorithm and a time-step length of $\delta t = 1.5 \cdot 10^{-3}$~ps. $\Gamma$-point sampling for integration within the first 
Brillouin zone was employed in all the AIMD simulations. 

Our finite-temperature simulations typically comprised long simulation times of $t_{total} \sim 100$~ps. For each material we 
typically ran an average of $3$ AIMD simulations at different temperatures within the range $300 \le T \le 1200$ K, considering 
both stoichiometric and non-stoichiometric compositions \cite{database}. Previous tests performed on the numerical bias stemming 
from the finite size of the simulation cell and duration of the molecular dynamics runs reported in work \cite{cazorla19} indicate 
that the adopted $N$ and $t_{total}$ values should provide reasonably well converged results for the ion diffusivity and 
vibrational density of states of SSE.

The mean squared displacement (MSD) was estimated with the formula:
\begin{equation}
{\rm MSD}(t)  = \frac{1}{N \left( N_{t} - n_{t} \right)} \cdot 
	        \sum_{i, j = 1}^{N, N_{t} - n_{t}} | {\bf r}_{i} (t_{j} + t) - {\bf r}_{i} (t_{j}) |^{2}~, 
\label{eq1}
\end{equation}
where ${\bf r}_{i}(t_{j})$ represents the position of the mobile ion $i$ at time $t_{j}$ ($= j \cdot \delta t$), $t$ a lag 
time, $n_{t} = t / \delta t$, $N$ the total number of mobile ions, and $N_{t}$ the total number of time steps 
(equivalent to $\sim 100$~ps). The maximum $n_{t}$ was chosen equal to $N_{t}/2$ (equivalent to $\sim 50$~ps) 
in order to accumulate enough statistics to reduce significantly the ${\rm MSD}(t)$ fluctuations at large $t$'s. The 
tracer diffusion coefficient, $D$, then was obtained based on the Einstein relation:
\begin{equation}
D =  \lim_{t \to \infty} \frac{{\rm MSD}(t)}{6t}~.  
\label{eq2}     
\end{equation}
In practice, we considered $0 < t \le 50$~ps and estimated $D$ by performing linear fits to the averaged ${\rm MSD}(t)$ 
obtained over the last $25$~ps. When taking into account many-ion correlations, the full diffusion coefficient was estimated 
by considering additional $i$--$j$ particle positions crossed terms in Eq.(\ref{eq1}) \cite{sasaki23}. 
\\

\textbf{Spatio-temporal correlation function.} The van Hove correlation function, $G (\Delta r, \Delta t)$, provides 
information on the spatio-temporal distribution of particles during a simulation. This two-dimensional function can be 
intuitively divided into a ``self'', $G_s$, and a ``distinct'', $G_d$, part. The former describes the displacements of 
a specific particle throughout time while the latter describes the relations of a particle with the rest, namely:
\begin{eqnarray}
	    G (r,t) & = & \frac{1}{N} \Bigg \langle \sum_{i,j=1}^{N} \delta{\left( r - \left| \mathbf{r}_i (t_{0} + t) 
	    - \mathbf{r}_j (t_{0}) \right| \right)} \Bigg \rangle \\ \nonumber
				   & = & G_{s} (r,t) + G_{d} (r,t)
\label{eq:vanHove}
\end{eqnarray}
where $\mathbf{r}$ represent the atomic positions, indices $i$ and $j$ run over all the mobile particles, $\delta (x)$ is the 
Dirac delta function, $t_{0}$ and arbitrary time, and averages are estimated over the total simulation time. The ``self'' and 
``distinct'' parts of the van Hove correlation function are then defined like:
\begin{eqnarray}
	G_{s} (r,t) & = &  \frac{1}{N} \Bigg \langle \sum_{i=1}^{N} \delta{\left(r - \left| \mathbf{r}_i (t_{0} + t) - 
	\mathbf{r}_i (t_{0}) \right| \right)} \Bigg \rangle \\ \nonumber
	G_{d} (r,t) & = & G (r,t) - G_{s} (r,t)~. 
\label{eq:self-dist}
\end{eqnarray}
\\

\textbf{IonDiff software.}~The freely available open-source python code \verb!IonDiff! \cite{iondiff} is based on an 
unsupervised k-means clustering algorithm (see next section for additional details). \verb!IonDiff! assigns a spatial 
point (i.e., centre of vibration) to every particle in the simulation supercell at each simulated time step. The centers 
of vibration then are compared with the stoichiometric equilibrium lattice so that (1)~ion-hopping events can be 
straightforwardly identified without the need of defining any arbitrary length or parameter, and (2)~metastable positions 
can be also readily determined. The residence time for a particular metastable position is estimated like the number of 
simulation steps associated with that location averaged over all the particles. The only required input files are two: 
(1)~one containing the positions of the particles throughout the whole simulation (e.g., XDATCAR file in the case of 
VASP calculations), and (2)~another detailing the length and number of time steps (e.g., INCAR file in the case of VASP 
calculations).
\\

\textbf{K-means clustering.}~The unsupervised algorithm devoted to identifying diffusive particles and their respective 
paths in molecular dynamics simulations is based on the k-means clustering approach. The implementation of the k-means 
clustering algorithm in the Scikit-learn python package \cite{scikit} was used in practice. The number of clusters at each 
time step, $K$, was selected based on the average silhouette method. In particular, the chosen $K$ corresponds to that which
maximises its average value over all possible $2 \le K$ cases (see main text). An arbitrary but reasonable confidence threshold 
value of $0.7$ was imposed for the silhouette coefficients, $S$ (Eq.\ref{eq:silhouette}). This means that if the maximum 
average silhouette coefficient amounted to less than $0.7$, the condition $K = 1$ was automatically imposed. 

Being $M_{I}$ the number of points in cluster $I$, with $M_{I} > 1$, the silhouette coefficient for a data point in that 
cluster, $i$, is mathematically defined like:
\begin{equation}
	S(i) = \frac{b(i) - a(i)}{\max{\left[ a(i), b(i) \right]}}~,
\label{eq:silhouette}
\end{equation}
where
\begin{eqnarray}
	a(i) = \frac{1}{M_{I} - 1} \sum_{j = 1, j \neq i}^{M_{I}} \| \mathbf{r}_j - \mathbf{r}_i \|^2 \\
	b(i) = \min_{J \neq I} \frac{1}{M_{J}} \sum_{j = 1}^{M_{J}} \| \mathbf{r}_j - \mathbf{r}_i \|^2~.
\label{eq:silhouette-2}
\end{eqnarray}
By proceeding in this manner, the similarity of a point within its own cluster and its dissimilarity with the others were
simultaneously optimized.
\\

\section*{Data Availability}
The data that support the findings of this study are available from the corresponding authors upon reasonable request.
\\

\section*{Acknowledgements}
C.C. acknowledges support from the Spanish Ministry of Science, Innovation and Universities under the fellowship 
RYC2018-024947-I, PID2020-112975GB-I00 and grant TED2021-130265B-C22. The authors thankfully acknowledge the CSIC under the 
``JAE Intro SOMdM 2021'' grant program and the computer resources at MareNostrum and the technical support provided by Barcelona 
Supercomputing Center (FI-1-0006, FI-2022-2-0003, FI-2023-1-0002, FI-2023-2-0004 and FI-2023-3-0004). R.R. acknowledges financial 
support from the MCIN/AEI/10.13039/501100011033 under Grant No. PID2020-119777GB-I00, the Severo Ochoa Centres of Excellence 
Program (CEX2019-000917-S) and the Generalitat de Catalunya under Grant No.2017SGR1506.
\\

\section*{Author Contributions}
C.C. conceived the study and planned the research. C.L. developed the analysis algorithms and applied them on a DFT-AIMD 
database previously generated by C.C., R.R. and C.L. Results were discussed by all the authors. The manuscript was written 
by C.L. and C.C. with substantial input from the rest of authors.
\\

\section*{Competing Interests}
The authors declare no competing interests.
\\

\end{document}